# A Methodology for Assessing the Environmental Effects Induced by ICT Services

## Part I: Single Services

Vlad C. Coroamă*
Department of Computer Science
ETH Zurich
Switzerland
vlad.coroama@inf.ethz.ch

Pernilla Bergmark*
Ericsson Research
Ericsson
Stockholm, Sweden
pernilla.bergmark@ericsson.com

Mattias Höjer
Department of Sustainable Development, Environmental Science and Engineering
KTH Royal Institute of Technology, Sweden
hojer@kth.se

Jens Malmodin
Ericsson Research
Ericsson
Stockholm, Sweden
jens.malmodin@ericsson.com

## ABSTRACT

Information and communication technologies (ICT) are increasingly seen as key enablers for climate change mitigation measures. They can make existing products and activities more efficient or substitute them altogether. Consequently, different initiatives have started to estimate the environmental effects of ICT services. Such assessments, however, lack scientific rigor and often rely on crude assumptions and methods, leading to inaccurate or even misleading results. The few methodological attempts that exist do not address several crucial aspects, and are thus insufficient to foster good assessment practice. Starting from such a high-level standard from the European Telecommunication Standardisation Institute (ETSI) and the International Telecommunication Union (ITU), this article identifies the shortcomings of existing methodologies and proposes solutions. It addresses several aspects for the assessment of single ICT services: the goal and scope definition (analyzing differences between ICT substitution and optimization, the time perspective of the assessment, the challenge of a hypothetical baseline for the situation without the ICT solution, and the differences between modelling and case studies) as well as the often-ignored influence of rebound effects and the difficult extrapolation from case studies to larger populations.

## 1 INTRODUCTION

To limit global warming to 1.5-2 degrees above preindustrial levels, humanity needs to halve its greenhouse gas (GHG) emissions every decade, with 2050 level to become only a small fraction of today´s emissions [1]. Information and communication technologies (ICT) are often envisioned as key enablers of such reductions throughout society. They can achieve this by, for example, substituting resource-intensive activities through ICT services – such as replacing conference travel through virtual connections that can entirely [2] or partially [3] virtualise conferences – or by making existing processes more efficient, for example different management services [4].

Sector-level claims referring to this potential have been put forward by the industry itself [5-11], but also by large international bodies such as the European Commission [12], OECD [13], and even the WWF [14-16]. Current methods and estimates, however, lack scientific rigour and often have to rely on crude assumptions and methods, as also acknowledged by these initiatives. Moreover, such estimates typically focus on services with expected benefits, ignoring those with possible negative effects. A new and accurate methodology thus needs to be developed in order to establish a more credible and consistent fact base.

The objective of this paper is thus to improve current assessment methods to provide for more rigorous assessments of the induced effects of ICT. Adding to established standards, this article undertakes a first step towards a comprehensive methodology for assessing the environmental effects induced by ICT services beyond their direct footprint. As a first step, it presents and categorizes the assessment challenges and reveals common flaws in existing industry claims. Based on this, the paper then proposes enhanced assessment principles for single services. The connected article, *A Methodology for Assessing the Environmental Effects Induced by ICT Services Part II: Multiple services and companies* [17], does the same for multiple services and companies.

The paper is structured as follows: Section 2 discusses the terminology and literature; Section 3 introduces the methodological basis and the contributions of the article; Section 4 analyses the challenges in assessing single ICT services, and proposes corresponding methodological enhancements. Section 5 discusses the limitations of our work and suggests directions for further research.

## 2 TERMINOLOGY

On the highest abstraction level, the literature distinguishes two main categories of environmental impacts for ICT:

- The impacts associated with the life cycle of ICT devices that, according to the life cycle assessment (LCA) standard [18], include raw materials acquisition, production, use, and end-of-life treatment. These impacts are also known as the direct (environmental) footprint. ICT's footprint has been addressed by numerous studies [19, 20].
- The other category contains a vast collection of subtler environmental effects induced by ICT infrastructure and devices. These effects range from the possible impact induced by an ICT service that provides virtual meetings (which might partly substitute physical meetings) to the long-term socio-economic consequences of ICT, such as economic growth or general behavioural changes [21].

While the direct footprint is always an environmental burden, effects in the second category can be environmentally either beneficial or detrimental. For both the footprint and for the second category of effects, different classifications and terminologies have been proposed: [22] considers all effects other than the footprint as indirect effects, while [23, 24] distinguish between first order (i.e., footprint), second order (effects induced by the use of ICT, such as efficiency gains or substitutions), and other effects (i.e., macroeconomic and behavioural consequences). Finally, [21] refers to ICT's own footprint as ICT infrastructure and devices, and distinguishes between three categories of further-reaching effects: applications, effects on economic growth and consumption patterns, and systemic effects on technology convergence and society. Without entering this terminological debate, the scope of our analysis coincides with the 'application' category from [21], as well as to 'second order effects' from [23, 24], taken together with the direct rebound effect from the 'other' category of the same standard, as discussed in Section 4.2.2. In the remainder of this paper, we refer to these effects as induced (environmental) effects, keeping in mind that this definition does not cover all the long-term behavioural and structural changes.

Concerns over ICT's environmental footprint, and visions of possible environmental benefits induced by ICT services, led to the emergence of two complementary research fields. The former domain is known as green ICT, while the latter under names such as green through ICT or ICT for sustainability (ICT4S) [25]. As these names show, much of the existing literature regarding the induced effects has focused exclusively on the positive effects induced by ICT, ignoring the possible negative outcomes. This is partly because many of the existing quantifications stem from the ICT industry itself. These positive assessments, too, used varying terminology. Several publications use the term abatement [6, 7, 9, 10]. As abatement refers to a decrease, it could describe both a reduction of ICT's own footprint (green ICT) and a reduction induced by ICT (green through ICT). To describe the latter, [26] put forward the term enablement, while [27] uses the term avoided emissions. In this paper, we adopt the more neutral term induced effect, indicating that such effect might be either positive or negative. When referring specifically to positive induced effects, we use synonymously the terms enablement or enabling effect, in line with the literature.

Some literature categorizes the mechanisms behind the enablement effect, without hints towards their quantification: [28] identified four such mechanisms through which ICT can induce GHG reductions, and more generally increase energy and resource efficiency: i) by substituting resource intensive activities through ICT services, ii) by making existing activities more efficient, iii) by intensifying the use of existing activities (slightly different than the previous in that existing activities do not become more efficient per usage, but can be used with increased frequency, such as enabling more trains per hour), and iv) by informing about existing consumption choices. In this paper, we exclude the special category 'iv) information', which involves indirect enabling mechanisms which are hard to quantify. We further group categories 'ii) increased efficiency' and 'iii) intensification' into one large category called *optimization*. All subsequent reflections and equations in this article refer to these two categories, *substitution* and *optimization*.

## 3 METHODOLOGICAL BASIS AND CONTRIBUTIONS

The main methodological reference for both the footprint and the induced effects of individual ICT services, is a standard jointly developed by ETSI [23] and ITU [24] – from now on referred to as the 'ETSI/ITU standard'.

This standard's main aim is to meticulously describe the direct life cycle footprint of ICT services, a topic outside our interests in this work. For the subtler induced effects, it only offers general guidance.

The standard abstractly defines the induced effects of one service through a comparative LCA between the footprint of an ICT service and the footprint of a reference activity representing the situation without ICT. This abstract guidance, however, ignores several issues encountered in practice. The most obvious one is that the comparison is inherently speculative, as both situations cannot coexist – at least one of them is necessarily hypothetical [29]. Moreover, most industry assessments refer to future potentials – in that case, not one but both LCAs are hypothetical. The necessary enhancement of the standard needs to address such issues. In this paper, we identify several issues for which further specification is needed. These are highlighted in Fig. 1, which reflects the structure of Fig. 24 of the ETSI/ITU standard, together with our proposed enhancements (in black).

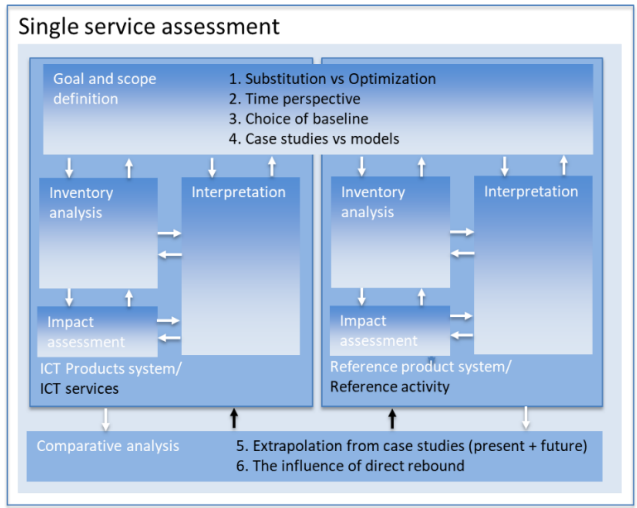

**Figure 1: This paper's suggested enhancements of the ETSI/ITU framework for the assessment of a single service are represented with numbered black text and symbols; unnumbered black text shows differences in terminology between this paper and the ETSI/ITU framework.**

Fig. 1 outlines the contributions of our work in relation to the ETSI/ITU framework. The figure show that the proposals refer to the initial goal and scoping phase as well as to the comparative analysis. For the actual assessment of the ICT service and of the reference activity no need for enhancing the standard was identified. For the scoping, with regard to a single service, we introduce the time perspective and the difference between substitution and optimization and their assessment implications. Furthermore, we address the complexity of defining a baseline for the reference activity, and, for the ICT service case, the use of case studies or models. For the comparative analysis, the extrapolation from case studies and the impact from direct rebound effects are discussed.

## 4 ASSESSING THE EFFECTS INDUCED BY ONE SERVICE

As defined by the ETSI/ITU standard and reflected in Fig. 1 above, the environmental effect of an ICT service is derived through a comparative LCA between the footprint of the reference activity without the ICT service, and the footprint of the service itself. As Fig. 1 shows, for this comparative LCA the following tasks need to be performed: i) definition of goal and scope of the study; ii) an LCA of the reference activity (including inventory analysis, impact assessment, and interpretation); iii) similarly, an LCA of the ICT service, and iv) a comparative analysis of the two.

As the comparative analysis is defined by subtracting the footprint of the ICT service from the footprint of the reference activity, the ETSI/ITU standard determines that a lower footprint of the substitute corresponds to a positive effect, i.e., an enablement. The individual LCAs from tasks ii) and iii) are well-defined through the ISO 14044 standard [18], and the supplementing ETSI/ITU standard. This paper focuses on tasks i) and iv), which are not sufficiently addressed by the standards.

### 4.1 Goal and Scope Definition

*4.1.1 Generalizing the ETSI/ITU basic principle: substitution versus optimization.* The joint ETSI/ITU standard proposes a simple formula to reflect the fact that the induced effect is the difference between the footprint of the reference activity and the footprint of the ICT service; using our terminology, it reads:

$$E(S_i|A_i) = FP(A_i) - FP(S_i) \qquad (1)$$

where $E(S_i|A_i)$ = the effect (in e.g. tonnes $CO_2e$) induced by ICT service $S_i$ for reference activity $A_i$; $FP(A_i)$ = the footprint of the reference activity $A_i$; and $FP(S_i)$ = the footprint of the ICT service $S_i$.

Eq. 1 implies that the standard focuses exclusively on substitutions, where a reference activity $A_i$ is substituted by an ICT service $S_i$. Optimizations would require a different formula, in which the reference activity $A_i$ still exists but in an optimized form $A_i'$. Moreover, for substitutions, the formula only applies to the subset of instances of $A_i$ that are substituted through instances of $S_i$. Smartphones, for instance, may reduce the purchase of standalone devices such as cameras, GPSs, etc., but currently do not fully substitute them. We thus define below an expanded equation that allows for partial substitutions (Eq. 2), by expanding the footprint function (FP) with a second argument, the set of usages (S). $FP(A_i, S)$ is then the footprint of all usages comprised in a set S, which equals $FP(A_i)$ when S includes all usages of $A_i$. For our continued discussion, we introduce two sets: M, which represents the usages of the reference activity that are modified by the ICT service; and N, which denotes the usages that are not affected.

Using these tools, we can now expand Eq. 1 to allow for partial substitutions. Better reflecting reality, the induced effect $E(S_i|A_i)$ is now computed as a difference between the footprint of the original activity for all initial instances M+N, $FP(A_i, M)+FP(A_i, N)$, and the footprint of the reference activity for the non-modified instances, $FP(A_i, N)$, as well as the footprint of the ICT service for the modified instances, $FP(S_i, M)$. As the footprint of the non-modified activities factors out, this equals to the difference between the footprint of the set M before modification, $FP(A_i, M)$, and its footprint after modification $FP(S_i, M)$:

$$E(S_i|A_i) = (FP(A_i, M)+FP(A_i, N)) - (FP(A_i, N) + FP(S_i, M))$$
$$= FP(A_i, M) - FP(S_i, M) \qquad (2)$$

Eq. 2 cannot be directly applied to an ICT service that does not substitute but optimizes a reference activity $A_i$. In such case, Eq. 2 needs to be expanded to also include the remaining footprint of the optimized reference activity $A_i'$, $FP(A_i',M)$:

$$E(S_i|A_i) = (FP(A_i, M)+FP(A_i, N)) - (FP(A_i, N) + FP(A_i',M) + FP(S_i, M)) = FP(A_i, M) - (FP(A_i', M) + FP(S_i, M)) \qquad (3)$$

Eq. 3 represents the general case. For substitutions, there is no optimized reference activity $A_i$', so $FP(A_i', M)$ equals zero, and the equation reduces to Eq. 2. From now on, we will thus use only Eq. 3 for services $S_i$ modifying (i.e., substituting or optimizing) reference activity $A_i$.

*4.1.2 Time perspective.* Past studies on the environmental effects of ICT services have not always clearly laid out whether they refer to a hypothetical ICT service not yet deployed or to an existing ICT service. They have also used (partly) inconsistent terminology. A future-oriented WWF study often uses the word potential [14]. A later WWF study [15] introduces the term *existing potentials*: yet untapped potentials that under certain assumptions can, in principle, be achieved today, with current technology and policies. This paper refers to this as a (circumstantial*)* *present potential*, as opposed to the former that might be called *future potential*, which would demand future changes such as further technological development or policy changes. Additionally, there is the *present* situation. To illustrate these concepts, the *present* effect may refer to the current number of usages of a telepresence service, the *present potential* to the effect that could occur if many more companies would start using the existing service, while the *future potential* would consider not only the potential usages of the future but also a development such as e.g. standardization of all telepresence platforms increasing their attractiveness and uptake. In this article, we clearly distinguish between *present*, *present potential*, and *future potentials*. Existing studies tend to postulate such potentials as positive; in this paper, however, we acknowledge that an ICT service might induce positive or negative effects [30]. This is also in line with [27], which advocates the avoidance of cherry-picking.

The effect of an ICT service substituting or optimizing a non-ICT reference activity is represented schematically in Fig. 2, where the service usage increases along the x-axis. The total environmental effect depends on both the number of usages, and the effect per usage (not explicitly shown in Fig. 2), both of which can be either measured or estimated. They are, in turn, functions of the parameters mentioned above. The service effect will vary over time depending on parameters such as technology development, and the legal, political and economic framework.

In Fig. 2, the aggregated present effect of an ICT service, and its dependence on the number of usages can be represented by a point P (present). The P-area represents the uncertainties referring to the number of usages and their aggregated effect, as well as to possibly ignored indirect effects (addressed in Section 4.2.2). If the service is not used at all, obviously it has no effect. As soon as the service is used, there will be positive or negative effects, represented by P. The present potential, PP, represents the aggregated effect from the maximum number of service usages possible today. In our example, it is a mildly positive one (as P as well); however, for other services it might be negative. Fig. 2 also denotes potential future usages and effects, the F-area. The future potential, F, depends on a variety of developments, and with an aggregated positive or negative effect. On the x-axis, F could in principle even be to the left of PP, if the usage of a service reduces over time. It should be noted that, while we refer to P, PP and F as points, they refer to a point in the graph of Fig. 2, not to a single point in time. A point in Fig. 2 stands for the number of usages and their aggregated effect during a particular time interval, typically a day, a month or a year. Finally, the CS-area in the figure represents effects related to case studies, which will be discussed in Section 4.1.4 below.

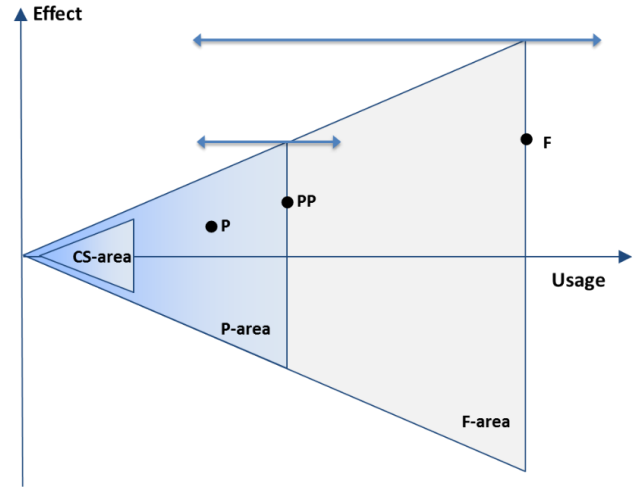

**Figure 2: Simple representation of an ICT service substituting an existing reference activity, showing the usage levels and environmental effects at present ('P-area'), in case studies ('CS-area') and in the future ('F-area'). PP represents the effect at maximum present usage. The present usage and effects, and those of case studies can, at least in principle, be measured, while PP and the projection-based future potential F are hypothetical points, established on specific assumptions.**

Studies should clearly state the object of their assessment, whether it is the effect of a case study (CS), the present (P), the present potential (PP), or some future potential (F). Thus, Eqs. 1-3 introduced earlier can be used to compute the induced effect for any of these four cases, i.e. for any point $X \in (CS, P, PP, F)$. To reflect this, we qualify Eq. 3 with an index that indicates the time perspective:

$E_X(S_i|A_i) =$
$(FP(A_i,M_X)+FP(A_i,N_X)) - (FP(A_i,N_X)+FP(A_i',M_X)+FP(S_i,M_X)) =$
$FP(A_i, M_X) - (FP(A_i', M_X) + FP(S_i, M_X))$ (4)

If the object of assessment is the future, we concur with [30] that an explicit distinction should be made between "data uncertainty (incomplete data about facts that could be known in principle) and future uncertainty (developments that are not determined and can be influenced in principle)". This distinction also shows that PP is hypothetical and thus conceptually more closely related to F than to P. Therefore, studies should make a clear distinction between data gaps of today´s achievements and assumptions for present and future potentials.

*4.1.3. Choice of baseline for the reference activity.* As soon as an ICT service is introduced, it is impossible to know how the reference activity would have developed without it. Still, its footprint needs to be estimated for the comparison as defined by the ETSI/ITU standard and reflected in Eqs. 1- 4. We call this hypothetical projection of the footprint of the reference activity the *baseline* and use it as a reference for estimating the effect induced by the ICT service.

The estimation of the induced environmental effect is highly dependent on the definition of this baseline [29]. Fig. 3 shows several possibilities of defining the baseline, as discussed by [32]: it can be fixed as a base value for the moment at which an ICT service was introduced (i), or the time of assessment (ii). It can, alternatively, be defined based on projections on how the situation without the ICT service would have evolved into the future (iii). In Fig. 3, (iii) illustrates a characteristic situation where the overall impact of the reference activity increases over time as the population and economy grow, i.e. more people can afford more activities. However, the reference activity might also have become more efficient over time, even without the ICT service. Uncertainties about efficiency gains, population and economic growth and other factors influence the shape of the projected baseline.

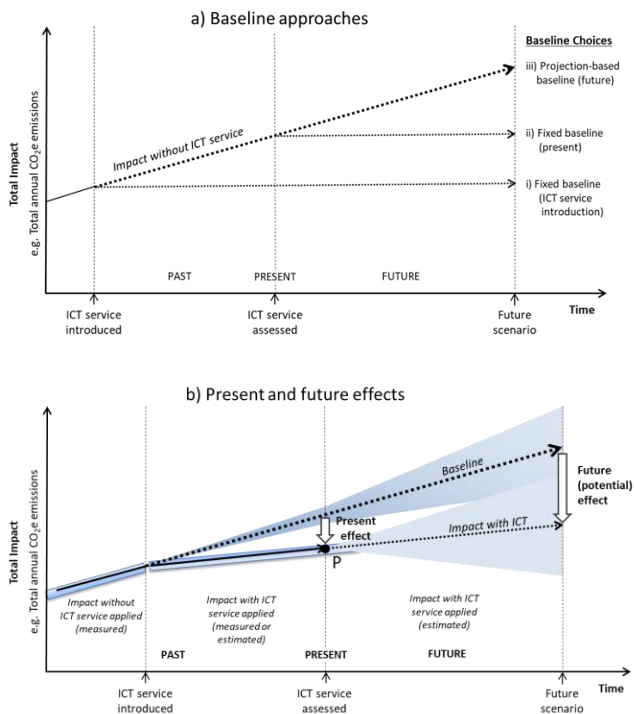

**Figure 3: Illustration of different approaches to define a baseline (3a), and the induced effect resulting from a projection-based baseline (3b). Fig. 3b shows the induced effect $E_X(S_i|A_i)$ of an ICT service $S_i$ already in use at the time of assessment. The (real) situation that includes the service is compared to a (hypothetical) situation without it represented by a projection-based baseline. Solid lines denote the actual situation measurable in principle; dotted lines indicate situations that did not happen (either as they represent a hypothetical situation without the introduced ICT service, or because they lie in the future); they can only be estimated. In both 4a and 4b, the y-axis devises the aggregated yearly effect of all instances of the service.**

Activities such as business travel [33] typically change over time. From the options introduced in Fig. 3a, we argue for the projection-based baseline (iii) in spite of the uncertainty of future extrapolation. This seems to best reflect reality, and is the option preferred by most of the literature, e.g. [32, 34], and also the IPCC [35]. Projection-based baselines should include efficiency gains due to expected technological progress, including the progress of ICT in general; the induced effect of a specific ICT service should only reflect the supplemental effect it induces. Such baselines should rely on relevant, recognized scenarios when they exist [36]. Fig. 3b thus shows such a baseline. The cone around it illustrates its uncertainties, and thus that different baseline scenarios might be equally valid.

Fig. 3b shows the situation when the service existed for some time already before the assessment of the present situation (P). Assuming that the service, from the moment of its introduction, did indeed induce a reduction of the environmental impact compared to the baseline, the total impact changes from the upper to the lower path in Fig. 3b. For simplicity, this path is represented linearly. Depending on the uptake of the service, possibly improved versions of the service, and other uncertainties, the gap to the baseline might widen faster, slower, or not at all into the future. This is represented by the lower cone of uncertainty in Fig. 3b.

*4.1.4 Case study versus models.* Assessing the overall effect $E_X(S_i|A_i)$ of an ICT service $S_i$ that substitutes or optimizes a reference activity is a bottom-up process which can either be modelled theoretically or derived from case studies [27]. With a theoretical approach, $E_X(S_i|A_i)$ could either be modelled directly, or an average per-usage effect $\tilde{e}$ could be estimated first, and subsequently multiplied by the number of service usages over a time period (day, month, year) as in Eq. 5. In this later case the average per-usage effect $\tilde{e}_{mod}(S_i|A_i)$ is modelled, and the overall effect $E_X(S_i|A_i)$ equals $\tilde{e}_{mod}$ multiplied by the amount of usages in the time interval under assessment. This number of usages equals by definition the cardinality of the set of modified instances M in point $X \in (P, PP, F)$, i.e., $|M_X|$:

$$E_X(S_i|A_i) = \tilde{e}_{mod}(S_i|A_i) * |M_X| \qquad (5)$$

Alternatively, case studies based on a limited set of usages can be deployed to understand the induced effect of ICT services. This principle of estimating a service's environmental effect based on scaled-up experiences from a pilot project is typical for services already in use, and deployed in both scientific [4] and industrial [9] contexts. The typical case study assessment procedure is to estimate the average induced effect per usage of the case study, $\tilde{e}_{CS}(S_i|A_i)$, and to extrapolate this average effect beyond the case study to all estimated usages in point $X \in (P, PP, F)$:

$$E_X(S_i|A_i) = \tilde{e}_{CS}(S_i|A_i) * |M_X| \qquad (6)$$

As this second path (i.e., case study and extrapolation) is usually taken by industry assessments, we focus our paper on this method. For case studies in general, covering both optimizations and substitutions, we introduce a function of the per-usage effect $e_j(S_i|A_i)$, which represents the effect of the modification of the jth instance of reference activity i ($A_i$), and is expressed as the difference between the footprint of the $j^{th}$ original reference activity, $fp_j(A_i)$, and the footprint of the $j^{th}$ usage of the ICT service $fp_j(S_i)$ together with the remaining footprint of the $j^{th}$ optimized $A_i'$, $fp_j(A_i')$, as shown in Eq. 7:

$$e_j(S_i|A_i) = fp_j(A_i) - ((fp_j(A_i') + fp_j(S_i))) \quad (7)$$

The sum of all per-usage effects yields the overall effect of the case study, $E_{CS}(S_i|A_i)$. Dividing this overall effect by the number of usages yields the average effect of the case study, $\tilde{e}_{CS}(S_i)$. These are reflected in Eqs. 8 and 9, respectively; $M_{CS}$ is hereby the set of modified activities within the case study. Unmodified activities within case studies are less common, but if they do exist, similarly to Eq. 4, they cancel out as they appear on both sides of the minus sign.

$$E_{CS}(S_i|A_i) = \sum_{j \in M_{CS}} e_j(S_i|A_i)$$
$$= \sum_{j \in M_{CS}} (fp_j(A_i) - (fp_j(A_i') + fp_j(S_i))) \quad (8)$$

$$\tilde{e}_{CS}(S_i|A_i) = E_{CS}(S_i|A_i) / |M_{CS}|$$
$$= (\sum_{j \in M_{CS}} (fp_j(A_i) - (fp_j(A_i') + fp_j(S_i)))) / |M_{CS}| \quad (9)$$

Eq. 9 shows how the average per-usage effect within a case study, $\tilde{e}_{CS}(S_i|A_i)$, is typically calculated. Combining it with the extrapolation to all usages in point $X \in (P, PP, F)$, as presented in Eq. 6, leads to:

$$E_X(S_i|A_i) = \tilde{e}_{CS}(S_i|A_i) * |M_X| = E_{CS}(S_i|A_i) * |M_X|/|M_{CS}| =$$
$$(\sum_{j \in M_{CS}} (fp_j(A_i) - (fp_j(A_i') + fp_j(S_i)))) * |M_X|/|M_{CS}| \quad (10)$$

## 4.2 Comparative analysis

The previous section more clearly laid out the goal and scope definition of the comparative analysis, insufficiently detailed in the high-level ETSI/ITU standard. After these needed additions, the current section addresses two challenges in performing the comparative analysis: the extrapolation from case studies, and the influence of rebound effects. Not acknowledging these two issues has been a frequent source of flaws in existing assessments, as further described in 4.1.1 and 4.1.2, respectively.

These two topics semantically belong to the comparative analysis (the lower box in Fig. 1), but both feed back into the assessment of the reference product system (in our terminology, reference activity) and/or the ICT product system (in our terminology, the ICT service). This shows that the very structure of Fig. 1 is in fact an over-simplification: the two product systems cannot be assessed separately and subsequently compared. In reality, an iterative approach is needed that switches between the individual assessments of the two product systems and their comparative analysis. Hence, at the very least, Fig. 1 should be extended by two arrows leading back from the comparative analysis to the analyzes of the two individual product systems. These are shown in Fig. 1 in black.

*4.2.1 Extrapolations from case studies.* The straightforward extrapolation in Eq. 10 reflects the typical way of estimating the effect of an ICT service via a case study. Although widely deployed, this method is appropriate only if two conditions are met: i) the assessment refers to the present effect, $E_P(S_i|A_i)$, and ii) the case study is based on a random sampling scheme. When these conditions apply, Eq. 10 can be used for assessing the present effect $E_P(S_i|A_i)$, and confidence intervals can be built using either classical statistical asymptotic theory or bootstrapping. Even in this simple case, however, the Hawthorne effect (of case study participants behaving differently due to the knowledge of being observed) might still skew the results.

For the present potential $E_{PP}(S_i)$ and any future potential effect $E_F(S_i|A_i)$, there are two additional sources of uncertainty:

- There is usually uncertainty regarding the potential adoption rate of the ICT service, i.e., its number of present and future potential usages, $|M_{PP}|$ and $|M_F|$, respectively. This is not always the case, however: for smart metering, for example, the potential usage pool is given by all the households of the assessed region.
- An additional uncertainty is the average effect per usage, $\tilde{e}_X(S_i|A_i)$. Many existing assessments implicitly set it on par with the average effect of the case study, $\tilde{e}_{CS}(S_i|A_i)$, but this can only be assumed when conditions i) and ii) above are met. We argue below why this assumption cannot be made for the assessment of $E_{PP}(S_i|A_i)$ or $E_F(S_i|A_i)$, or even $E_P(S_i|A_i)$ if the case study sample was not randomly selected among all present usages.

Regarding the second uncertainty, even if there is a good estimate for today's average service impact, the average impact per usage may change over time or between user groups, or, as identified by [36], due to contextual factors. As further groups of potential users are involved, the question is whether the current ones can be regarded as representative for all, as also discussed in [4]. Different biases might render any extrapolation to additional groups challenging, such as geography, income, etc.

For assessments of the present, a bias of particular importance is whether today's users participate in an early pilot on a voluntary basis. If so, the overrepresentation of early adopters in the sample might not be in line with a wider usage and an extrapolation would suffer from 'volunteer biased sampling' [37]. The influence of the volunteer bias on the result can be quite sizeable – in a meta-study of smart meter reports, [38] found that small, volunteer-based case studies report savings of up to 15-20%, while broader studies with random sampling only report savings of around 2%.

The uncertainty regarding the number of usages is also multifaceted. There are novelty effects like more intense usage of a service initially, or lower usage if the usefulness of the service depends on the number of users applying it. The frequency of usage for services such as videoconferencing essentially depends on market penetration. If many users have access to videoconferencing rooms, it is likely that the use of the system will increase. As technology progresses or individual users progressively become more familiarized with an ICT service, the frequency or efficiency of use might also increase. A traffic-aware navigation system, on the other hand, when used by only a few users, might bring substantial savings through suggesting alternative routes. Nevertheless, with increasing uptake of the service, the clotting of alternate routes might lead to decreasing per-usage savings.

As a pragmatic way to account for these uncertainties and biases, we propose a first, simplistic approach, in which extrapolations from case studies are multiplied by a conservatively chosen ex-trapolation coefficient, which will bring the estimates more in line with realistic expectations. With this extrapolation coefficient $k_X$, which models the extrapolation from $\tilde{e}_{CS}(S_i|A_i)$ to $\tilde{e}_X(S_i|A_i)$ for $X \in \{P, PP, F\}$, the rudimentary Eq. 6 for extrapolation from case studies becomes

$$E_X(S_i|A_i) = k_X * \tilde{e}_{CS}(S_i|A_i) * |M_X| \quad (11)$$

Accordingly, Eq. 10 becomes

$$E_X(S_i|A_i) = k_X * E_{CS}(S_i|A_i) * |M_X| / |M_{CS}| =$$
$$k_X * (\sum_{j \in M_{CS}} (fp_j(A_i) - (fp_j(A_i') + fp_j(S_i)))) * |M_X|/|M_{CS}| \quad (12)$$

If the case study is statistically relevant for an assessment of P, $k_X$ is 1, and Eq. 12 reduces to Eq. 10, with $E_X(S_i|A_i)$ expanded according to Eq. 8. In any other case, we recommend a conservative $k_X$ to be used. If the users have been chosen on a volunteer basis, $k_X$ could be chosen as low as 0.1-0.2, to rather err on the conservative side. This initial approach, which copes in a pragmatic way with the lack of data for more rigorous methods, represents a substantial improvement over the often-encountered naïve extrapolations from volunteer case studies to entire populations. Moving forward, more rigorous approaches are needed, based on statistical methods, or on the generalization of long-term empiric insights on how initial effects evolved over time for different services. When statistical data is lacking, we concur with [27, 36] that the use of alternative scenarios and sensitivity analyses are important tools to make the uncertainty visible.

*4.2.2 The influence of direct rebound.* When efficiency gains make a product or service more affordable, classical economic theory teaches that demand for it increases, all other things equal. The per-unit efficiency gains may be partially or entirely offset, or even overcompensated, by this increased demand. This seeming paradox was coined *rebound effect* by [39]. Its approach relies on a single-service model; i.e., there are no repercussions from this one service to the rest of the economy. Today, the narrow phenomenon described by Khazzoom is thus usually called *direct rebound effect*, and there is an entire class of wider effects named *indirect rebound effects*. Definitions of this broader understanding are given by e.g. [22, 40-42], how rebound effects are related to digitization is discussed in [43].

Here, we only focus on the direct rebound, which is immediately relevant to the environmental assessment of ICT services. We show both how its existence affects the equations introduced thus far, and discuss how ignoring the rebound effect can lead to serious overestimates of the induced effect of the ICT service.

Fig. 4a illustrates the rebound effect and its theoretical influence on the assessment in the substitution case. It presents how the environmental effect resulting from the introduction of an ICT service $S_i$ (e.g., telepresence) that partly substitutes an original reference activity $A_i$ (e.g., travel) affects the systemic environmental effect. As discussed earlier, the original instances of $A_i$ are now divided into a set of non-modified activities $N_X$ and a set of modified activities $M_X$. Before considering rebound, in accordance with Eq. 2 the effect of substituting $S_i$ for $A_i$ is $FP(A_i, M_X) - FP(S_i, M_X)$; the dashed rectangle in the figure. The availability of the cheaper and easily accessible service can, however, lead to a rebound effect represented by the light green bar in Fig. 4a. The set R of rebound instances has no correspondent in the activities $A_i$ (e.g., the attendees of those videoconferences would not have held a corresponding physical meeting), and their footprint is $FP(S_i, R_X)$. Expanding thus the basic Eq. 2 to include the rebound effect of substitutions, leads to a modified equation, corresponding to subtracting the light green rectangle from the dashed one in Fig. 4a:

$$E_X(S_i|A_i) = FP(A_i, M_X) - FP(S_i, M_X + R_X) \quad (13)$$

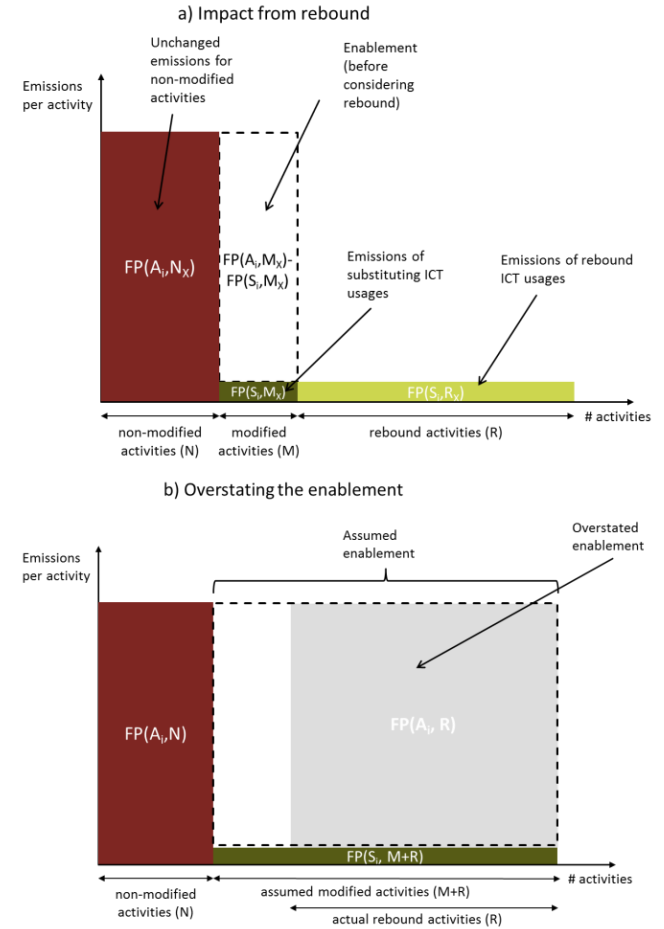

**Figure 4: Schematic illustration of the impact from rebound activities (5a), and an often-encountered error in existing assessments (5b). In Fig. 4a, an ICT service $S_i$ (dark green) substitutes a reference activity $A_i$, inducing – before subtracting rebound – the enablement represented by the dashed rectangle. The induced rebound effect (light green), however, reduces this assumed enablement. Eq. 13 reflects the situation in Fig. 4a, while Eq. 14 shows the general case of modifications (not shown graphically). Fig. 4b shows the effect when all usages of $S_i$ are considered to be modified activities although some are in fact additional usages, due to rebound. This can lead to a large overstatement of the positive effect of the ICT modification (grey area, described by Eq. 15).**

To not be overly cluttered, Fig. 4a presents only substitutions and not the general case of optimizations. For an optimized service, a further horizontal bar of the same length as FP($S_i$, $M_X$ + $R_X$) would represent the footprint of the modified reference activity FP($A_i$', $M_X$ + $R_X$), and the general Eq. 4 becomes:

$$E_X(S_i|A_i) = FP(A_i, M_X) - (FP(A_i', M_X + R_X) + FP(S_i, M_X + R_X)) \quad (14)$$

The rebound effect thus counteracts the environmental benefits of an ICT service. Although the number of rebound instances might be large, however, the overall effect can still be beneficial as the footprint of the rebound instances of the ICT substitute is often small, or even negligible, compared to that of the substituted reference activity (for example, of videoconferencing as compared to flying).

The main rebound-related source of mistakes, however, is not ignoring its footprint, but a different, subtler one. It stems from the fact that once the service is in place, it is in most cases an ontological uncertainty to distinguish between modified activities and rebound activities, i.e., between the sets M and R. The influence this uncertainty can have on assessments is shown in Fig. 4b: existing assessments often wrongly assume that all existing instances of service $S_i$ modified previously existing instances of reference activity $A_i$, ignoring that many of them may in fact be rebound due to the attractiveness of $S_i$. Assuming this naïve view leads to an overstated assessed effect of

$$E_X(S_i|A_i) = FP(A_i, M_X + \mathbf{R_X}) - (FP(A_i', M_X + R_X) + FP(S_i, M_X + R_X)) \quad (15)$$

The overestimate from Eq. 15 equals the grey-coloured area in Fig. 4b, FP($A_i$, $R_X$). In particular if the rebound is extremely large (up to orders of magnitude as compared to the original reference activity), ignoring it can lead to substantial overestimates of the environmental effect of an ICT service. This is a subtle effect because it does not directly relate to the size of the rebound itself, but to the assumption that the modified activities would have been similarly-sized, had the modifying service not existed. As [44] puts it, the estimated enabling effect "can always be increased by defining a pessimistic baseline".

To illustrate this error in practice, a telecom company computed an enabling effect of 10 megatons (Mt) $CO_2e$ for its ICT services [45]. Among several services, the largest part of this reduction came from cellphone conversations that replaced other means of communication. The paper claims that 5.5% of the cellphone calls in 2010 replaced physical meetings, resulting in considerable GHG emission reductions. It is, however, impossible to exactly determine whether any particular instance of the substitute replaced the reference activity or not. Rebound effects might be responsible for 1% or for 99.9% of the usage, as compared to the reference activity.

It is even more important to consider rebound effects for future assessments. The same [45] further asserts that the enablement almost tripled from 10 Mt $CO_2e$ in 2010 to 29 Mt $CO_2e$ in 2013. For 2013, the study used the same 5.5% as percentage of calls replacing travel. As in 2013 the absolute number of calls over the cellular network grew abruptly, the same percentage of 5.5% represents a much higher absolute value. This seems a second fallacy related to rebound effects and extrapolation. Even if this percentage was a correct estimate for 2010, the later increased service usage is most likely a consequence of the attractiveness of the service, i.e., a rebound effect, as it does not seem plausible that the need for face-to-face meetings would increase drastically between 2010 and 2013. This example shows how the influence of rebound effects may become more significant over time.

While helpful for conceptualizing rebound effects and their influence on the result, the equations in this section do not help in assessing their magnitude. The equations are merely a way of raising awareness how the failure to account for rebound might lead to a significant overstatement of the enablement. As rebound effects might have a crucial influence on the result, well-founded and possibly conservative assumptions seem key to a robust assessment. It does not seem a solid scientific basis having users speculate among rebound and original usage, as in [3, 45], which both use a survey as the sole basis of information. When the usage intensity of an ICT service is used as a basis for computing the environmental benefits of the service, overstatements may occur, intentionally or not. A possible solution could be to consider the usage intensity of the service (i.e., M+N+R) as one variable among many. Other criteria should be used for validation. When assessing the effect of high-quality videoconferencing on travel, for example, the usage intensity should not be the only, neither the determinant, factor in computing the induced effect. Other attributes should be considered, such as past levels of travel for the particular company under assessment, changes in travel costs and number of employees, rate of travel by similar companies not deploying the service, specific travel demand, broader societal trends, etc. Rebound effects tend to become more relevant as time passes. Future assessments should thus be even more cautious in avoiding ICT service's usage intensity as a sole basis for the assessment.

*4.2.3 Bringing extrapolations from case studies and rebound together.* To refine the comparative analysis of the high-level ETSI/ITU framework for the induced environmental effect of ICT services (see Fig. 1), we discussed here the extrapolation from case studies as well as the direct rebound effect and its influence on the assessment. Combining these insights to account for rebound within case studies, the calculation of the case study effect in Eq. 8 needs to be expanded by a second term that represents the footprint of the rebound usages, corresponding to the set $R_{CS}$, yielding Eq. 16:

$$E_{CS}(S_i|A_i) = \sum_{j \in M_{CS}}(fp_j(A_i) - (fp_j(A_i') + fp_j(S_i))) - \sum_{j \in R_{CS}}(fp_j(S_i) + fp_j(A_i')) = \sum_{j \in M_{CS}}(fp_j(A_i)) - \sum_{j \in (M_{CS}, R_{CS})}(fp_j(S_i) + fp_j(A_i')) \quad (16)$$

Accordingly, accounting for rebound effects in the computation of the environmental effect of a service $S_i$ via a case study, changes Eq. 12 to

$$E_X(S_i|A_i) = k_X * (\sum_{j \in M_{CS}}(fp_j(A_i)) - \sum_{j \in (M_{CS}, R_{CS})}(fp_j(S_i) + fp_j(A_i'))) * |M_X| / |M_{CS}| \quad (17)$$

## 5 DISCUSSION AND FUTURE RESEARCH

Section 5.1 discusses the relevance of the methodology enhancements put forward in this paper. Section 5.2 addresses areas for future research. For those aspects we identified the issue but, given the breadth of the analysis and the complexity of the topics, we only proposed a first solution, which is often simple and pragmatic. Below, we discuss how these solutions could be further developed.

### 5.1 Is there a need for enhanced assessment methods for the induced effects of ICT?

Through their capacity to optimize various activities or substitute them altogether, ICT services could be important contributors towards reducing greenhouse gas emissions but could also have adverse effects. An increasing number of ICT companies, industry associations, and even public initiatives have started to estimate induced effects of ICT. As this paper describes, current estimations use heterogeneous approaches, and often ignore the complexity of the topic, neglect negative effects and often overlook rebound effects, which leads to overstatements of the induced effect.

Still, the domain itself remains highly relevant, and an assessment methodology which reflects the complexity of the induced effects of ICT services is needed. Enabling companies to rigorously assess and report their positive contributions could, moreover, provide more accurate knowledge and support green funding initiatives and trading schemes to more accurately assess services, thus helping to optimize the use of ICT services for the necessary decarbonization of society.

### 5.2 Hypothetical baseline and uncertain extrapolation

The assessment guidelines presented here are based on the comparison between a situation with an ICT service and a reference activity represented by the baseline. Such baseline, however, needs to be conceptually meaningful – the situation without the service must be a conceivable alternative. There is a point in time when a service becomes part of everyday life and mainstream technology. In the vocabulary of transition management, it is the moment when a "technological niche" has become part of the current "sociotechnical regime" [46]. When such situation arrives, companies should better avoid making further claims for the induced effect of the service; how to assess its advent is subject of further, multidisciplinary research.

Eqs. 11-12 in Section 4.2 suggested the usage of a conservatively chosen extrapolation coefficient to avoid the overestimates often encountered in literature when extrapolating from case studies. This coefficient should reflect the quality of a case study, in particular its sample size, user categories covered (e.g., early adopters vs a broader range), whether the average effect is likely to correlate positively, negatively, or behave indifferent to uptake, and whether the enablement effect induced by the service might be synergistically linked to the usage of other foreseeable ICT or non-ICT activities. This basic approach requires further study and refinement, possibly using propensity scores or other statistical methods.

## 6 CONCLUSION

Starting from the ETSI/ITU framework outlined in Fig. 1 and Eq. 1, this paper identified challenges for understanding the environmental effects induced by ICT services through substitution or optimization of reference activities. Beyond identifying common flaws in existing assessments, this article put forward solutions as a step towards establishing a more rigorous and comprehensive methodology for assessing the induced effects of one ICT service.

For the assessment of individual ICT services, areas identified and addressed include substituting versus optimizing ICT services, the time perspective (P, PP and F), choice of baseline for the reference activity, assessment via the tricky extrapolation of case studies, and the impact of direct rebound effects (in particular how they could be misinterpreted as substitutions). In a next step, the assessment principles should cover a set of services and the allocation to companies. We provide an attempt in that direction in *A Methodology for Assessing the Environmental Effects Induced by ICT Services Part II: Multiple services and companies* [17].

In addition to a theoretical analysis, principles are proposed for addressing most of these issues when performing assessments. Thus, though not providing a cookbook recipe, our contribution both addresses the complexities, but also offers guidance to practitioners. Future work should seek to further develop the parts of our approach where we only offer high-level guidance or preliminary workarounds. Moreover, the proposed methodology should be tested on several real cases – an exercise that in all its needed complexity would have gone far beyond the scope of the current paper. Nevertheless, the approach established in this article should certainly improve current practices and enable more accurate identification, classification and ranking of low-carbon ICT services.

## ACKNOWLEDGEMENTS

The authors wish to thank Dr. Åsa Moberg for general feedback on earlier versions of this article. This research was financed by the organizations of the authors and with co-funding for the KTH-part from Vinnova, Sweden's innovation agency.